\documentclass{eptcs}

\usepackage{breakurl}
\usepackage[fleqn]{amsmath}
\RequirePackage[T1]{fontenc}
\RequirePackage[utf8x]{inputenc}
\RequirePackage{ucs}
\makeatletter
\@ifundefined{lhs2tex.lhs2tex.sty.read}%
  {\@namedef{lhs2tex.lhs2tex.sty.read}{}%
   \newcommand\SkipToFmtEnd{}%
   \newcommand\EndFmtInput{}%
   \long\def\SkipToFmtEnd#1\EndFmtInput{}%
  }\SkipToFmtEnd

\newcommand\ReadOnlyOnce[1]{\@ifundefined{#1}{\@namedef{#1}{}}\SkipToFmtEnd}
\usepackage{amstext}
\usepackage{amssymb}
\usepackage{stmaryrd}
\DeclareFontFamily{OT1}{cmtex}{}
\DeclareFontShape{OT1}{cmtex}{m}{n}
  {<5><6><7><8>cmtex8
   <9>cmtex9
   <10><10.95><12><14.4><17.28><20.74><24.88>cmtex10}{}
\DeclareFontShape{OT1}{cmtex}{m}{it}
  {<-> ssub * cmtt/m/it}{}

\DeclareFontShape{OT1}{cmtt}{bx}{n}
  {<5><6><7><8>cmtt8
   <9>cmbtt9
   <10><10.95><12><14.4><17.28><20.74><24.88>cmbtt10}{}
\DeclareFontShape{OT1}{cmtex}{bx}{n}
  {<-> ssub * cmtt/bx/n}{}
\newcommand{\Conid}[1]{\mathit{#1}}
\newcommand{\Varid}[1]{\mathit{#1}}
\newcommand{\anonymous}{\kern0.06em \vbox{\hrule\@width.5em}}

\usepackage{polytable}

\@ifundefined{mathindent}%
  {\newdimen\mathindent\mathindent\leftmargini}%
  {}%

\def\resethooks{%
  \global\let\SaveRestoreHook\empty
  \global\let\ColumnHook\empty}
\newcommand*{\savecolumns}[1][default]%
  {\g@addto@macro\SaveRestoreHook{\savecolumns[#1]}}
\newcommand*{\restorecolumns}[1][default]%
  {\g@addto@macro\SaveRestoreHook{\restorecolumns[#1]}}
\newcommand*{\aligncolumn}[2]%
  {\g@addto@macro\ColumnHook{\column{#1}{#2}}}

\resethooks

\newcommand{\onelinecommentchars}{\quad-{}- }
\newcommand{\commentbeginchars}{\enskip\{-}
\newcommand{\commentendchars}{-\}\enskip}

\newcommand{\visiblecomments}{%
  \let\onelinecomment=\onelinecommentchars
  \let\commentbegin=\commentbeginchars
  \let\commentend=\commentendchars}

\newcommand{\invisiblecomments}{%
  \let\onelinecomment=\empty
  \let\commentbegin=\empty
  \let\commentend=\empty}

\visiblecomments

\newlength{\blanklineskip}
\newcommand{\hsindent}[1]{\quad}% default is fixed indentation
\let\hspre\empty
\let\hspost\empty

\EndFmtInput
\makeatother
\ReadOnlyOnce{polycode.fmt}%
\makeatletter

\newcommand{\hsnewpar}[1]%
  {{\parskip=0pt\parindent=0pt\par\vskip #1\noindent}}

\newcommand{\hscodestyle}{}

\newcommand{\sethscode}[1]%
  {\expandafter\let\expandafter\hscode\csname #1\endcsname
   \expandafter\let\expandafter\endhscode\csname end#1\endcsname}

  {\par\noindent
   \advance\leftskip\mathindent
   \hscodestyle
   \let\\=\@normalcr
   \let\hspre\(\let\hspost\)%
   \pboxed}%
  {\endpboxed\)%
   \par\noindent
   \ignorespacesafterend}

  {\hsnewpar\abovedisplayskip
   \advance\leftskip\mathindent
   \hscodestyle
   \let\hspre\(\let\hspost\)%
   \pboxed}%
  {\endpboxed%
   \hsnewpar\belowdisplayskip
   \ignorespacesafterend}

  {\hsnewpar\abovedisplayskip
   \advance\leftskip\mathindent
   \hscodestyle
   \let\\=\@normalcr
   \(\pboxed}%
  {\endpboxed\)%
   \hsnewpar\belowdisplayskip
   \ignorespacesafterend}

\newcommand{\plainhs}{\sethscode{plainhscode}}

\plainhs

  {\hsnewpar\abovedisplayskip
   \advance\leftskip\mathindent
   \hscodestyle
   \let\\=\@normalcr
   \(\parray}%
  {\endparray\)%
   \hsnewpar\belowdisplayskip
   \ignorespacesafterend}

  {\parray}{\endparray}

  {\(\parray}{\endparray\)}

\def\codeframewidth{\arrayrulewidth}
\RequirePackage{calc}

  {\parskip=\abovedisplayskip\par\noindent
   \hscodestyle
   \arrayrulewidth=\codeframewidth
   \tabular{@{}|p{\linewidth-2\arraycolsep-2\arrayrulewidth-2pt}|@{}}%
   \hline\framedhslinecorrect\\{-1.5ex}%
   \let\endoflinesave=\\
   \let\\=\@normalcr
   \(\pboxed}%
  {\endpboxed\)%
   \framedhslinecorrect\endoflinesave{.5ex}\hline
   \endtabular
   \parskip=\belowdisplayskip\par\noindent
   \ignorespacesafterend}

\newcommand{\framedhslinecorrect}[2]%
  {#1[#2]}

  {\(\def\column##1##2{}%
   \let\>\undefined\let\<\undefined\let\\\undefined
   \newcommand\>[1][]{}\newcommand\<[1][]{}\newcommand\\[1][]{}%
   \def\fromto##1##2##3{##3}%
   }{\) }%

  {\let\orighscode=\hscode
   \let\origendhscode=\endhscode
   \def\endhscode{\def\hscode{\endgroup\def\@currenvir{hscode}\\}\begingroup}
   \orighscode\def\hscode{\endgroup\def\@currenvir{hscode}}}%
  {\origendhscode
   \global\let\hscode=\orighscode
   \global\let\endhscode=\origendhscode}%

\makeatother
\EndFmtInput

\def\commentbegin{\{{-}~}

\usepackage{todonotes}
\usepackage{tikz}
\usepackage{tikz-cd}
\def\doubleequals{\mathrel{\unitlength 0.01em
  \begin{picture}(78,40)
    \put(7,34){\line(1,0){25}} \put(45,34){\line(1,0){25}}
    \put(7,14){\line(1,0){25}} \put(45,14){\line(1,0){25}}
  \end{picture}}}
\title{Examples and Results from a BSc-level Course on\\ Domain Specific Languages of Mathematics}

\author{
  Patrik Jansson
  \institute{Chalmers Univ. of Technology}
  \email{\quad patrikj@chalmers.se}
\and
  Sólrún Halla Einarsdóttir
  \institute{Chalmers Univ. of Technology}
  \email{\quad slrn@chalmers.se}
\and
  Cezar Ionescu
  \institute{University of Oxford}
  \email{\quad cezar.ionescu@conted.ox.ac.uk}
}

\def\titlerunning{Examples from DSLs of Mathematics}
\def\authorrunning{P. Jansson \& S. H. Einarsdóttir \& C. Ionescu}
\newcommand{\event}{7th International Workshop on Trends in Functional Programming in Education, TFPIE 2018}

\def\commentbegin{\{{-}~}

\begin{document}

\maketitle

\begin{abstract}
  At the workshop on Trends in Functional Programming in Education (TFPIE) in
  2015 Ionescu and Jansson presented the approach underlying the ``Domain
  Specific Languages of Mathematics'' (DSLsofMath) course even before the first
  course instance.
  We were then encouraged to come back to present our experience and the student
  results.
  Now, three years later, we have seen three groups of learners attend the
  course, and the first two groups have also continued on to take challenging
  courses in the subsequent year.
  In this paper we present three examples from the course material to set the
  scene, and we present an evaluation of the student results showing
  improvements in the pass rates and grades in later courses.
\end{abstract}


\section{Introduction}

For the last few years we have been working on the border between
education and functional programming research under the common heading
of ``Domain Specific Languages of Mathematics''
(\href{https://github.com/DSLsofMath/}{DSLsofMath}).
This activity started from a desire to improve the mathematical
education of computer scientists and the computer science education of
mathematicians.
In 2014 Ionescu and Jansson applied for a pedagogical project grant to
develop a new BSc level course, and from 2016 on the course has been
offered to students at Chalmers and University of Gothenburg.

At the workshop on Trends in Functional Programming in Education
(TFPIE) in 2015 Ionescu and Jansson
\cite{DBLP:journals/corr/IonescuJ16} presented the approach underlying
the DSLsofMath course even before the first course instance.
We were then encouraged to come back to present our experience and the
student results.
Now, three years later, we have seen three groups of learners attend
the course, and the first two groups have also continued on to take
mathematically challenging compulsory courses in the subsequent year.

The course focus is on types and specifications and on the syntax and semantics
of domain specific languages used as tools for thinking
(for more details see appendix \ref{app:learningoutcomes}).
In this paper we present three examples from the course material to
set the scene, and we present an evaluation of the student results.

The DSLsofMath activity has also lead to other developments not
covered in this paper: presentations at TFPIE 2015, the Workshop on
Domain Specific Languages Design and Implementation (DSLDI 2015), IFIP
Working Group 2.1 on Algorithmic Languages and Calculi meeting in
2015, and two BSc thesis projects (one in 2016 about Transforms,
Signals, and Systems \cite{JonssonTSLwithDLS2016} and one in 2018
called ``Learn You a Physics''\cite{LearnYouaPhysicsBSc2018} --- see
appendix \ref{app:LearnYouAPhysics}).

% WG2.1: 2015 http://www.cse.chalmers.se/~patrikj/talks/WG2.1_Goteborg_Jansson_Ionescu_DSLsofMath.pdf

\section{Static checking: Scope and Types in Mathematics}

The DSLsofMath lecture notes \cite{JanssonIonescuDSLsofMathCourse}
have evolved from raw text notes for the first instance to 152 pages
of PDF generated from literate Haskell + LaTeX sources for the third
instance.
To give the reader a feeling for the course contents, we will show two
smaller and, in the next section, one larger examples from the lecture notes.
The two smaller examples are limits and derivatives.

% In several places the book contains an indented quote of a definition
% or paragraph from a mathematical textbook, followed by detailed
% analysis of that quote.
% %
% The aim is to improve the reader's skills in understanding, modelling,
% and implementing mathematical text.

In many of the chapters we start from a textbook definition and ``tease it
apart'' to identify parameters, types, and to help the students
understand exactly what it means.
When we first presented the course
\cite{DBLP:journals/corr/IonescuJ16}, we stressed the importance of
syntax and semantics, types and specifications.
As the material developed we noticed that variable binding (and scope)
is also an important (and in mathematical texts often implicit)
ingredient.
In our first example here, limits, we show the students that an
innocent-looking ``if A then B'' can actually implicitly bind one of
the names occurring in A.

\subsection{Case 1: Scoping Mathematics: The limit of a function}
\label{sec:LimitOfFunction}

This case is from Chapter two of the DSLsofMath lecture notes which
talks about the definition (from Adams \& Essex
\cite{adams2010calculus}) of the limit of a function of type \ensuremath{\mathbb{R}\to \mathbb{R}}:

\begin{quote}
  We say that \(f(x)\) \textbf{approaches the limit} \(L\) as \(x\)
  \textbf{approaches} \(a\), and we write
  \[\lim_{x\to a} f(x) = L,\]
  if the following condition is satisfied:\\
  for every number \(\epsilon > 0\) there exists a number
  \(\delta > 0\), possibly depending on \(\epsilon\), such that if
  \ensuremath{\mathrm{0}\mathbin{<}\lvert{}\Varid{x}\mathbin{-}\Varid{a}\rvert{}\mathbin{<}\delta}, then \(x\) belongs to the domain of \(f\)
  and
  \begin{hscode}\SaveRestoreHook
\column{B}{@{}>{\hspre}l<{\hspost}@{}}%
\column{5}{@{}>{\hspre}l<{\hspost}@{}}%
\column{E}{@{}>{\hspre}l<{\hspost}@{}}%
\>[5]{}\lvert{}\Varid{f}\;(\Varid{x})\mathbin{-}\Conid{L}\rvert{}\mathbin{<}\epsilon.{}\<[E]%
\ColumnHook
\end{hscode}\resethooks

\end{quote}

\noindent
The \ensuremath{\Varid{lim}} notation has four components: a variable name \ensuremath{\Varid{x}}, a point
\ensuremath{\Varid{a}}, an expression \(f(x)\), and the limit \ensuremath{\Conid{L}}.
The variable name and the expression can be combined into just the
function \ensuremath{\Varid{f}} and this leaves us with three essential components: \ensuremath{\Varid{a}}, \ensuremath{\Varid{f}},
and \ensuremath{\Conid{L}}.
Thus, \ensuremath{\Varid{lim}} can be seen as a ternary (3-argument) predicate which is
satisfied if the limit of \ensuremath{\Varid{f}} exists at \ensuremath{\Varid{a}} and equals \ensuremath{\Conid{L}}.
If we apply our logic toolbox we can define \ensuremath{\Varid{lim}} starting something like this:
\begin{hscode}\SaveRestoreHook
\column{B}{@{}>{\hspre}l<{\hspost}@{}}%
\column{12}{@{}>{\hspre}c<{\hspost}@{}}%
\column{12E}{@{}l@{}}%
\column{15}{@{}>{\hspre}l<{\hspost}@{}}%
\column{E}{@{}>{\hspre}l<{\hspost}@{}}%
\>[B]{}\Varid{lim}\;\Varid{a}\;\Varid{f}\;\Conid{L}{}\<[12]%
\>[12]{}\mathrel{=}{}\<[12E]%
\>[15]{}\forall\ \epsilon\mathbin{>}\mathrm{0}.\ \ \exists\ \delta\mathbin{>}\mathrm{0}.\ \ \Conid{P}\;\epsilon\;\delta{}\<[E]%
\ColumnHook
\end{hscode}\resethooks
It is often useful to introduce a local name (like \ensuremath{\Conid{P}} here) to help
break the definition down into more manageable parts.
If we now naively translate the last part we get this ``definition''
for \ensuremath{\Conid{P}}:
\begin{hscode}\SaveRestoreHook
\column{B}{@{}>{\hspre}l<{\hspost}@{}}%
\column{14}{@{}>{\hspre}l<{\hspost}@{}}%
\column{21}{@{}>{\hspre}l<{\hspost}@{}}%
\column{88}{@{}>{\hspre}l<{\hspost}@{}}%
\column{E}{@{}>{\hspre}l<{\hspost}@{}}%
\>[B]{}\quad\;{}\<[14]%
\>[14]{}\mathbf{where}\;{}\<[21]%
\>[21]{}\Conid{P}\;\epsilon\;\delta\mathrel{=}(\mathrm{0}\mathbin{<}\lvert{}\Varid{x}\mathbin{-}\Varid{a}\rvert{}\mathbin{<}\delta)\Rightarrow (\Varid{x}\in \Conid{Dom}\;\Varid{f}{}\<[88]%
\>[88]{}\mathrel{\wedge}\lvert{}\Varid{f}\;\Varid{x}\mathbin{-}\Conid{L}\rvert{}\mathbin{<}\epsilon)){}\<[E]%
\ColumnHook
\end{hscode}\resethooks
Note that there is a scoping problem: we have \ensuremath{\Varid{f}}, \ensuremath{\Varid{a}}, and \ensuremath{\Conid{L}} from
the ``call'' to \ensuremath{\Varid{lim}} and we have \ensuremath{\epsilon} and \ensuremath{\delta} from the two
quantifiers, but where did \ensuremath{\Varid{x}} come from?
It turns out that the formulation ``if \ldots then \ldots'' hides a
quantifier that binds \ensuremath{\Varid{x}}.
Thus we get this definition:
\begin{hscode}\SaveRestoreHook
\column{B}{@{}>{\hspre}l<{\hspost}@{}}%
\column{3}{@{}>{\hspre}l<{\hspost}@{}}%
\column{10}{@{}>{\hspre}l<{\hspost}@{}}%
\column{12}{@{}>{\hspre}c<{\hspost}@{}}%
\column{12E}{@{}l@{}}%
\column{15}{@{}>{\hspre}l<{\hspost}@{}}%
\column{31}{@{}>{\hspre}l<{\hspost}@{}}%
\column{92}{@{}>{\hspre}l<{\hspost}@{}}%
\column{109}{@{}>{\hspre}l<{\hspost}@{}}%
\column{E}{@{}>{\hspre}l<{\hspost}@{}}%
\>[B]{}\Varid{lim}\;\Varid{a}\;\Varid{f}\;\Conid{L}{}\<[12]%
\>[12]{}\mathrel{=}{}\<[12E]%
\>[15]{}\forall\ \epsilon\mathbin{>}\mathrm{0}.\ \ \exists\ \delta\mathbin{>}\mathrm{0}.\ \ \Conid{P}\;\epsilon\;\delta{}\<[E]%
\\[\blanklineskip]%
\>[B]{}\hsindent{3}{}\<[3]%
\>[3]{}\mathbf{where}\;{}\<[10]%
\>[10]{}\Conid{P}\;\epsilon\;\delta\mathrel{=}{}\<[31]%
\>[31]{}\forall\ \Varid{x}.\ \ \Conid{Q}\;\epsilon\;\delta\;\Varid{x}{}\<[E]%
\\[\blanklineskip]%
\>[10]{}\Conid{Q}\;\epsilon\;\delta\;\Varid{x}\mathrel{=}{}\<[31]%
\>[31]{}(\mathrm{0}\mathbin{<}\lvert{}\Varid{x}\mathbin{-}\Varid{a}\rvert{}\mathbin{<}\delta)\Rightarrow {}\<[92]%
\>[92]{}(\Varid{x}\in \Conid{Dom}\;\Varid{f}{}\<[109]%
\>[109]{}\mathrel{\wedge}\lvert{}\Varid{f}\;\Varid{x}\mathbin{-}\Conid{L}\rvert{}\mathbin{<}\epsilon){}\<[E]%
\ColumnHook
\end{hscode}\resethooks
The predicate \ensuremath{\Varid{lim}} can be shown to be a partial function of two
arguments, \ensuremath{\Varid{f}} and \ensuremath{\Varid{a}}.
This means that each function \ensuremath{\Varid{f}} can have \emph{at most} one limit
\ensuremath{\Conid{L}} at a point \ensuremath{\Varid{a}}.
(This is not evident from the definition and proving it is a good
exercise.)

\subsection{Case 2: Typing Mathematics: derivative of a function}

The lecture notes include some other material in between this example and the
previous one, but here we
jump directly to dissecting one of the classical definitions of the
derivative (also from \cite{adams2010calculus}).
We now assume limits exist and use \ensuremath{\Varid{lim}} as a function from \ensuremath{\Varid{a}} and \ensuremath{\Varid{f}} to \ensuremath{\Conid{L}}.

\begin{quote}
  The \textbf{derivative} of a function \ensuremath{\Varid{f}} is another function \ensuremath{\Varid{f'}} defined by
  \[
    f'(x) = \lim_{h \to 0} \frac{f(x+h) - f(x)}{h}
  \]
  at all points \ensuremath{\Varid{x}} for which the limit exists (i.e., is a finite real
  number). If \(f'(x)\) exists, we say that \ensuremath{\Varid{f}} is \textbf{differentiable}
  at \ensuremath{\Varid{x}}.
\end{quote}
We can start by assigning types to the expressions in the definition.
Let's write \ensuremath{\Conid{X}} for the domain of \ensuremath{\Varid{f}} so that we have \ensuremath{\Varid{f}\mathbin{:}\Conid{X}\to \mathbb{R}}
and \ensuremath{\Conid{X}\;\subseteq\;\mathbb{R}} (or, equivalently, \ensuremath{\Conid{X}\mathbin{:}\mathcal{P}\;\mathbb{R}}).
If we denote with \ensuremath{\Conid{Y}} the subset of \ensuremath{\Conid{X}} for which \ensuremath{\Varid{f}} is
differentiable we get \ensuremath{\Varid{f'}\mathbin{:}\Conid{Y}\to \mathbb{R}}.
Thus, the operation which maps \ensuremath{\Varid{f}} to \ensuremath{\Varid{f'}} has type \linebreak \ensuremath{(\Conid{X}\to \mathbb{R})\to (\Conid{Y}\to \mathbb{R})}.
Unfortunately, the only notation for this operation given (implicitly)
in the definition is a postfix prime.
To make it easier to see we use a prefix \ensuremath{\Conid{D}} instead and we can
thus write \ensuremath{\Conid{D}\mathbin{:}(\Conid{X}\to \mathbb{R})\to (\Conid{Y}\to \mathbb{R})}.
We will often assume that \ensuremath{\Conid{X}\mathrel{=}\Conid{Y}} so that we can can see \ensuremath{\Conid{D}} as
preserving the type of its argument.

Now, with the type of \ensuremath{\Conid{D}} sorted out, we can turn to the actual
definition of the function \ensuremath{\Conid{D}\;\Varid{f}}.
The definition is given for a fixed (but arbitrary) \ensuremath{\Varid{x}}.
The \ensuremath{\Varid{lim}} expression is using the (anonymous) function \ensuremath{\Varid{g}\;\Varid{h}\mathrel{=}\frac{\Varid{f}\;(\Varid{x}\mathbin{+}\Varid{h})\mathbin{-}\Varid{f}\;\Varid{x}}{\Varid{h}}} and that the limit of \ensuremath{\Varid{g}} is taken at \ensuremath{\mathrm{0}}.
Note that \ensuremath{\Varid{g}} is defined in the scope of \ensuremath{\Varid{x}} and that its definition
uses \ensuremath{\Varid{x}} so it can be seen as having \ensuremath{\Varid{x}} as an implicit, first
argument.
To be more explicit we write \ensuremath{\varphi\;\Varid{x}\;\Varid{h}\mathrel{=}\frac{\Varid{f}\;(\Varid{x}\mathbin{+}\Varid{h})\mathbin{-}\Varid{f}\;\Varid{x}}{\Varid{h}}} and take
the limit of \ensuremath{\varphi\;\Varid{x}} at 0.
So, to sum up, \ensuremath{\Conid{D}\;\Varid{f}\;\Varid{x}\mathrel{=}\Varid{lim}\;\mathrm{0}\;(\varphi\;\Varid{x})}.
We could go one step further by noting that \ensuremath{\Varid{f}} is in the scope of \ensuremath{\varphi} and used in its definition.
Thus the function \ensuremath{\psi\;\Varid{f}\;\Varid{x}\;\Varid{h}\mathrel{=}\varphi\;\Varid{x}\;\Varid{h}}, or \ensuremath{\psi\;\Varid{f}\mathrel{=}\varphi}, is used.
With this notation we obtain a point-free definition that can come in
handy:
\ensuremath{\Conid{D}\;\Varid{f}\mathrel{=}\Varid{lim}\;\mathrm{0}\mathbin{\circ}\psi\;\Varid{f}}.
To sum up, here are the steps again, now with typed helpers:

\newsavebox{\diagramD}
\savebox{\diagramD}{%
\begin{tikzcd}
         \pgfmatrixnextcell \arrow[dl, "\ensuremath{\Conid{D}\;\Varid{f}}", swap] \arrow[d, "\ensuremath{\psi\;\Varid{f}}"] \ensuremath{\mathbb{R}} \\
  \ensuremath{\mathbb{R}} \pgfmatrixnextcell \arrow[l, "\ensuremath{\Varid{lim}\;\mathrm{0}}"] \ensuremath{(\mathbb{R}\to \mathbb{R})}
\end{tikzcd}%
}
\begin{hscode}\SaveRestoreHook
\column{B}{@{}>{\hspre}l<{\hspost}@{}}%
\column{3}{@{}>{\hspre}l<{\hspost}@{}}%
\column{10}{@{}>{\hspre}l<{\hspost}@{}}%
\column{27}{@{}>{\hspre}l<{\hspost}@{}}%
\column{34}{@{}>{\hspre}l<{\hspost}@{}}%
\column{39}{@{}>{\hspre}l<{\hspost}@{}}%
\column{44}{@{}>{\hspre}l<{\hspost}@{}}%
\column{47}{@{}>{\hspre}l<{\hspost}@{}}%
\column{87}{@{}>{\hspre}l<{\hspost}@{}}%
\column{92}{@{}>{\hspre}l<{\hspost}@{}}%
\column{113}{@{}>{\hspre}l<{\hspost}@{}}%
\column{122}{@{}>{\hspre}l<{\hspost}@{}}%
\column{E}{@{}>{\hspre}l<{\hspost}@{}}%
\>[3]{}\Conid{D}\;\Varid{f}\;\Varid{x}{}\<[10]%
\>[10]{}\mathrel{=}\Varid{lim}\;\mathrm{0}\;\Varid{g}\;{}\<[27]%
\>[27]{}\mathbf{where}\;{}\<[44]%
\>[44]{}\Varid{g}\;{}\<[47]%
\>[47]{}\Varid{h}\mathrel{=}\frac{\Varid{f}\;(\Varid{x}\mathbin{+}\Varid{h})\mathbin{-}\Varid{f}\;\Varid{x}}{\Varid{h}};\quad{}\<[87]%
\>[87]{}\Varid{g}{}\<[92]%
\>[92]{}\mathbin{:}{}\<[122]%
\>[122]{}\mathbb{R}\to \mathbb{R}{}\<[E]%
\\
\>[3]{}\Conid{D}\;\Varid{f}\;\Varid{x}{}\<[10]%
\>[10]{}\mathrel{=}\Varid{lim}\;\mathrm{0}\;(\varphi\;\Varid{x})\;{}\<[27]%
\>[27]{}\mathbf{where}\;{}\<[39]%
\>[39]{}\varphi\;{}\<[44]%
\>[44]{}\Varid{x}\;{}\<[47]%
\>[47]{}\Varid{h}\mathrel{=}\frac{\Varid{f}\;(\Varid{x}\mathbin{+}\Varid{h})\mathbin{-}\Varid{f}\;\Varid{x}}{\Varid{h}};{}\<[87]%
\>[87]{}\varphi{}\<[92]%
\>[92]{}\mathbin{:}{}\<[113]%
\>[113]{}\mathbb{R}\to {}\<[122]%
\>[122]{}\mathbb{R}\to \mathbb{R}{}\<[E]%
\\
\>[3]{}\Conid{D}\;\Varid{f}{}\<[10]%
\>[10]{}\mathrel{=}\Varid{lim}\;\mathrm{0}\mathbin{\circ}\psi\;\Varid{f}\;{}\<[27]%
\>[27]{}\mathbf{where}\;{}\<[34]%
\>[34]{}\psi\;{}\<[39]%
\>[39]{}\Varid{f}\;{}\<[44]%
\>[44]{}\Varid{x}\;{}\<[47]%
\>[47]{}\Varid{h}\mathrel{=}\frac{\Varid{f}\;(\Varid{x}\mathbin{+}\Varid{h})\mathbin{-}\Varid{f}\;\Varid{x}}{\Varid{h}};{}\<[87]%
\>[87]{}\psi{}\<[92]%
\>[92]{}\mathbin{:}(\mathbb{R}\to \mathbb{R})\to {}\<[113]%
\>[113]{}\mathbb{R}\to {}\<[122]%
\>[122]{}\mathbb{R}\to \mathbb{R}\usebox{\diagramD}{}\<[E]%
\ColumnHook
\end{hscode}\resethooks

The key here is that we name, type, and specify the operation of
computing the derivative (of a one-argument function).
This operation is used quite a bit in the rest of the lecture notes,
but here are just a few examples to get used to the notation.

\begin{hscode}\SaveRestoreHook
\column{B}{@{}>{\hspre}l<{\hspost}@{}}%
\column{3}{@{}>{\hspre}l<{\hspost}@{}}%
\column{13}{@{}>{\hspre}c<{\hspost}@{}}%
\column{13E}{@{}l@{}}%
\column{16}{@{}>{\hspre}l<{\hspost}@{}}%
\column{E}{@{}>{\hspre}l<{\hspost}@{}}%
\>[3]{}\Conid{D}\mathbin{:}(\mathbb{R}\to \mathbb{R})\to (\mathbb{R}\to \mathbb{R}){}\<[E]%
\\[\blanklineskip]%
\>[3]{}\Varid{sq}\;\Varid{x}{}\<[13]%
\>[13]{}\mathrel{=}{}\<[13E]%
\>[16]{}\Varid{x}\mathbin{\hat{}}\mathrm{2}{}\<[E]%
\\
\>[3]{}\Varid{double}\;\Varid{x}{}\<[13]%
\>[13]{}\mathrel{=}{}\<[13E]%
\>[16]{}\mathrm{2}\mathbin{*}\Varid{x}{}\<[E]%
\\
\>[3]{}\Varid{c}_{2}\;\Varid{x}{}\<[13]%
\>[13]{}\mathrel{=}{}\<[13E]%
\>[16]{}\mathrm{2}{}\<[E]%
\ColumnHook
\end{hscode}\resethooks
Then we have the following equalities:
\begin{hscode}\SaveRestoreHook
\column{B}{@{}>{\hspre}l<{\hspost}@{}}%
\column{3}{@{}>{\hspre}l<{\hspost}@{}}%
\column{9}{@{}>{\hspre}c<{\hspost}@{}}%
\column{9E}{@{}l@{}}%
\column{13}{@{}>{\hspre}l<{\hspost}@{}}%
\column{20}{@{}>{\hspre}l<{\hspost}@{}}%
\column{E}{@{}>{\hspre}l<{\hspost}@{}}%
\>[3]{}\Varid{sq'}{}\<[9]%
\>[9]{}\doubleequals{}\<[9E]%
\>[13]{}\Conid{D}\;\Varid{sq}{}\<[20]%
\>[20]{}\doubleequals\Conid{D}\;(\lambda \Varid{x}\to \Varid{x}\mathbin{\hat{}}\mathrm{2})\doubleequals\Conid{D}\;({}\mathbin{\hat{}}\mathrm{2})\doubleequals(\mathrm{2}\mathbin{*})\doubleequals\Varid{double}{}\<[E]%
\\
\>[3]{}\Varid{sq''}{}\<[9]%
\>[9]{}\doubleequals{}\<[9E]%
\>[13]{}\Conid{D}\;\Varid{sq'}{}\<[20]%
\>[20]{}\doubleequals\Conid{D}\;\Varid{double}\doubleequals\Varid{c}_{2}\doubleequals\Varid{const}\;\mathrm{2}{}\<[E]%
\ColumnHook
\end{hscode}\resethooks

What we cannot do at this stage is to actually \emph{implement} \ensuremath{\Conid{D}} in
Haskell.
If we only have a function \ensuremath{\Varid{f}\mathbin{:}\mathbb{R}\to \mathbb{R}} as a ``black box'' we
cannot really compute the actual derivative \ensuremath{\Varid{f'}\mathbin{:}\mathbb{R}\to \mathbb{R}}, only
numerical approximations.
But if we also have access to the ``source code'' of \ensuremath{\Varid{f}}, then we can
apply the usual rules we have learnt in calculus.

\section{Type inference and understanding: Lagrangian case study}
\label{sec:Lagrangian}

The lecture notes goes on through several other definitions related to
derivatives, including partial derivatives \(D_i = ∂\!f/∂x_i\) of type
\((ℝⁿ → ℝ) → (ℝⁿ → ℝ)\).
As an example, we can define \ensuremath{\Conid{D₃}} in terms of \ensuremath{\Conid{D}}:
\begin{hscode}\SaveRestoreHook
\column{B}{@{}>{\hspre}l<{\hspost}@{}}%
\column{3}{@{}>{\hspre}l<{\hspost}@{}}%
\column{31}{@{}>{\hspre}l<{\hspost}@{}}%
\column{E}{@{}>{\hspre}l<{\hspost}@{}}%
\>[B]{}\Conid{D₃}\;\Varid{f}\;(\Varid{x}_{1},\Varid{x}_{2},\Varid{x}_{3})\mathrel{=}\Conid{D}\;\Varid{g}\;\Varid{x}_{3}{}\<[E]%
\\
\>[B]{}\hsindent{3}{}\<[3]%
\>[3]{}\mathbf{where}\;\Varid{g}\;\Varid{x}\mathrel{=}\Varid{f}\;(\Varid{x}_{1},\Varid{x}_{2},\Varid{x}){}\<[31]%
\>[31]{}\mbox{\onelinecomment  \ensuremath{\Varid{g}\mathbin{:}\Conid{ℝ}\mathbin{→}\Conid{ℝ}} keeps \ensuremath{\Varid{x}_{1}} and \ensuremath{\Varid{x}_{2}} constant}{}\<[E]%
\ColumnHook
\end{hscode}\resethooks

Our third case study from the lecture notes is the analysis of
Lagrangian equations, also studied in Sussman and Wisdom 2013
\cite{sussman2013functional} in their prologue on ``Programming and
Understanding''.

\begin{quote}
  A mechanical system is described by a Lagrangian function of the
  system state (time, coordinates, and velocities).
  A motion of the system is described by a path that gives the
  coordinates for each moment of time.
  A path is allowed if and only if it satisfies the Lagrange
  equations.
  Traditionally, the Lagrange equations are written

\[\frac{d}{dt} \frac{∂L}{∂\dot{q}} - \frac{∂L}{∂q} = 0\]

What could this expression possibly mean?

\end{quote}

\noindent
In order to answer Sussman and Wisdom's question, we start by inferring the types of the elements involved:

\begin{enumerate}
\item The use of notation for ``partial derivative'', \(∂L / ∂q\), suggests
that \ensuremath{\Conid{L}} is a function of at least a pair of arguments:
\begin{hscode}\SaveRestoreHook
\column{B}{@{}>{\hspre}l<{\hspost}@{}}%
\column{3}{@{}>{\hspre}l<{\hspost}@{}}%
\column{18}{@{}>{\hspre}l<{\hspost}@{}}%
\column{E}{@{}>{\hspre}l<{\hspost}@{}}%
\>[3]{}\Conid{L}\mathbin{:}\Conid{ℝⁱ}\mathbin{→}\Conid{ℝ},{}\<[18]%
\>[18]{}\Varid{i}\mathbin{≥}\mathrm{2}{}\<[E]%
\ColumnHook
\end{hscode}\resethooks

This is consistent with the description: ``Lagrangian function of the
system state (time, coordinates, and velocities)''.
So, if we let ``coordinates'' be just one coordinate, we can take \ensuremath{\Varid{i}\mathrel{=}\mathrm{3}}:
\begin{hscode}\SaveRestoreHook
\column{B}{@{}>{\hspre}l<{\hspost}@{}}%
\column{3}{@{}>{\hspre}l<{\hspost}@{}}%
\column{E}{@{}>{\hspre}l<{\hspost}@{}}%
\>[3]{}\Conid{L}\mathbin{:}\Conid{ℝ³}\mathbin{→}\Conid{ℝ}{}\<[E]%
\ColumnHook
\end{hscode}\resethooks
The ``system state'' here is a triple (of type \ensuremath{\Conid{S}\mathrel{=}(\Conid{T},\Conid{Q},\Conid{V})\mathrel{=}\Conid{ℝ³}})
and we can call the three components \ensuremath{\Varid{t}\mathbin{:}\Conid{T}} for time, \ensuremath{\Varid{q}\mathbin{:}\Conid{Q}} for
coordinate, and \ensuremath{\Varid{v}\mathbin{:}\Conid{V}} for velocity.
(We use \ensuremath{\Conid{T}\mathrel{=}\Conid{Q}\mathrel{=}\Conid{V}\mathrel{=}\Conid{ℝ}} in this example but it can help the reading to
remember the different uses of \ensuremath{\Conid{ℝ}}.)

\item Looking again at the same derivative, \(∂L / ∂q\) suggests that
  \(q\) is the name of a real variable, one of the three arguments to
  \(L\).
  In the context, which we do not have, we would expect to find
  somewhere the definition of the Lagrangian as
  \begin{hscode}\SaveRestoreHook
\column{B}{@{}>{\hspre}l<{\hspost}@{}}%
\column{5}{@{}>{\hspre}l<{\hspost}@{}}%
\column{8}{@{}>{\hspre}c<{\hspost}@{}}%
\column{8E}{@{}l@{}}%
\column{11}{@{}>{\hspre}l<{\hspost}@{}}%
\column{22}{@{}>{\hspre}c<{\hspost}@{}}%
\column{22E}{@{}l@{}}%
\column{26}{@{}>{\hspre}c<{\hspost}@{}}%
\column{26E}{@{}l@{}}%
\column{E}{@{}>{\hspre}l<{\hspost}@{}}%
\>[5]{}\Conid{L}{}\<[8]%
\>[8]{}\mathbin{:}{}\<[8E]%
\>[11]{}(\Conid{T},\Conid{Q},\Conid{V}){}\<[22]%
\>[22]{}\to {}\<[22E]%
\>[26]{}\Conid{ℝ}{}\<[26E]%
\\
\>[5]{}\Conid{L}\;{}\<[11]%
\>[11]{}(\Varid{t},\Varid{q},\Varid{v}){}\<[22]%
\>[22]{}\mathrel{=}{}\<[22E]%
\>[26]{}\mathbin{...}{}\<[26E]%
\ColumnHook
\end{hscode}\resethooks

\item therefore, \(∂L / ∂q\) should also be a function of the same
  triple of arguments:

  \begin{hscode}\SaveRestoreHook
\column{B}{@{}>{\hspre}l<{\hspost}@{}}%
\column{5}{@{}>{\hspre}l<{\hspost}@{}}%
\column{E}{@{}>{\hspre}l<{\hspost}@{}}%
\>[5]{}(\mathbin{∂}\Conid{L}\mathbin{/}\mathbin{∂}\Varid{q})\mathbin{:}(\Conid{T},\Conid{Q},\Conid{V})\to \Conid{ℝ}{}\<[E]%
\ColumnHook
\end{hscode}\resethooks

  It follows that the equation expresses a relation between
  \emph{functions}, therefore the \(0\) on the right-hand side is
  \emph{not} the real number \(0\), but rather the constant function
  \ensuremath{\Varid{const}\;\mathrm{0}}:

  \begin{hscode}\SaveRestoreHook
\column{B}{@{}>{\hspre}l<{\hspost}@{}}%
\column{5}{@{}>{\hspre}l<{\hspost}@{}}%
\column{14}{@{}>{\hspre}c<{\hspost}@{}}%
\column{14E}{@{}l@{}}%
\column{17}{@{}>{\hspre}l<{\hspost}@{}}%
\column{28}{@{}>{\hspre}c<{\hspost}@{}}%
\column{28E}{@{}l@{}}%
\column{31}{@{}>{\hspre}l<{\hspost}@{}}%
\column{32}{@{}>{\hspre}l<{\hspost}@{}}%
\column{E}{@{}>{\hspre}l<{\hspost}@{}}%
\>[5]{}\Varid{const}\;\mathrm{0}{}\<[14]%
\>[14]{}\mathbin{:}{}\<[14E]%
\>[17]{}(\Conid{T},\Conid{Q},\Conid{V}){}\<[28]%
\>[28]{}\mathbin{→}{}\<[28E]%
\>[31]{}\Conid{ℝ}{}\<[E]%
\\
\>[5]{}\Varid{const}\;\mathrm{0}\;{}\<[17]%
\>[17]{}(\Varid{t},\Varid{q},\Varid{v}){}\<[28]%
\>[28]{}\mathrel{=}{}\<[28E]%
\>[32]{}\mathrm{0}{}\<[E]%
\ColumnHook
\end{hscode}\resethooks

\item We now have a problem: \ensuremath{\Varid{d}\mathbin{/}\Varid{dt}} can only be applied to functions
  of \emph{one} real argument \ensuremath{\Varid{t}}, and the result is a function of one
  real argument:

\begin{hscode}\SaveRestoreHook
\column{B}{@{}>{\hspre}l<{\hspost}@{}}%
\column{5}{@{}>{\hspre}l<{\hspost}@{}}%
\column{38}{@{}>{\hspre}c<{\hspost}@{}}%
\column{38E}{@{}l@{}}%
\column{41}{@{}>{\hspre}l<{\hspost}@{}}%
\column{E}{@{}>{\hspre}l<{\hspost}@{}}%
\>[5]{}(\Varid{d}\mathbin{/}\Varid{dt})\,(\mathbin{∂}\Conid{L}\mathbin{/}\mathbin{∂}\dot{q}){}\<[38]%
\>[38]{}\mathbin{:}{}\<[38E]%
\>[41]{}\Conid{T}\mathbin{→}\Conid{ℝ}{}\<[E]%
\ColumnHook
\end{hscode}\resethooks

Since we subtract from this the function \(∂L / ∂q\), it follows that
this, too, must be of type \ensuremath{\Conid{T}\to \Conid{ℝ}}.
But we already typed it as \ensuremath{(\Conid{T},\Conid{Q},\Conid{V})\mathbin{→}\Conid{ℝ}}, contradiction!
\label{item:L:contra}

\item The expression \(∂L / ∂\dot{q}\) appears to also be malformed.
  We would expect a variable name where we find \(\dot{q}\), but
  \(\dot{q}\) is the same as \(dq / dt\), a function.

\item Looking back at the description above, we see that the only
  immediate candidate for an application of \(d/dt\) is ``a path that
  gives the coordinates for each moment of time''.
  Thus, the path is a function of time, let us say
  \begin{hscode}\SaveRestoreHook
\column{B}{@{}>{\hspre}l<{\hspost}@{}}%
\column{5}{@{}>{\hspre}l<{\hspost}@{}}%
\column{8}{@{}>{\hspre}c<{\hspost}@{}}%
\column{8E}{@{}l@{}}%
\column{11}{@{}>{\hspre}l<{\hspost}@{}}%
\column{18}{@{}>{\hspre}l<{\hspost}@{}}%
\column{E}{@{}>{\hspre}l<{\hspost}@{}}%
\>[5]{}\Varid{w}{}\<[8]%
\>[8]{}\mathbin{:}{}\<[8E]%
\>[11]{}\Conid{T}\mathbin{→}\Conid{Q}{}\<[18]%
\>[18]{}\mbox{\onelinecomment  with \ensuremath{\Conid{T}\mathrel{=}\Conid{ℝ}} for time and \ensuremath{\Conid{Q}\mathrel{=}\Conid{ℝ}} for coordinates (\ensuremath{\Varid{q}\mathbin{:}\Conid{Q}})}{}\<[E]%
\ColumnHook
\end{hscode}\resethooks

  We can now guess that the use of the plural form ``equations'' might
  have something to do with the use of ``coordinates''.
  In an \ensuremath{\Varid{n}}-dimensional space, a position is given by \ensuremath{\Varid{n}}
  coordinates.
  A path would then be a function
  \begin{hscode}\SaveRestoreHook
\column{B}{@{}>{\hspre}l<{\hspost}@{}}%
\column{5}{@{}>{\hspre}l<{\hspost}@{}}%
\column{8}{@{}>{\hspre}c<{\hspost}@{}}%
\column{8E}{@{}l@{}}%
\column{11}{@{}>{\hspre}l<{\hspost}@{}}%
\column{18}{@{}>{\hspre}l<{\hspost}@{}}%
\column{E}{@{}>{\hspre}l<{\hspost}@{}}%
\>[5]{}\Varid{w}{}\<[8]%
\>[8]{}\mathbin{:}{}\<[8E]%
\>[11]{}\Conid{T}\mathbin{→}\Conid{Q}{}\<[18]%
\>[18]{}\mbox{\onelinecomment  with \ensuremath{\Conid{Q}\mathrel{=}\Conid{ℝⁿ}}}{}\<[E]%
\ColumnHook
\end{hscode}\resethooks
  which is equivalent to \ensuremath{\Varid{n}} functions of type \ensuremath{\Conid{T}\mathbin{→}\Conid{ℝ}}, each computing
  one coordinate as a function of time.
  We would then have an equation for each of them.
  We will use \ensuremath{\Varid{n}\mathrel{=}\mathrm{1}} for the rest of this example.

\item Now that we have a path, the coordinates at any time are given
  by the path.
  And as the time derivative of a coordinate is a velocity, we can
  actually compute the trajectory of the full system state \ensuremath{(\Conid{T},\Conid{Q},\Conid{V})}
  starting from just the path.
  \begin{hscode}\SaveRestoreHook
\column{B}{@{}>{\hspre}l<{\hspost}@{}}%
\column{5}{@{}>{\hspre}l<{\hspost}@{}}%
\column{8}{@{}>{\hspre}c<{\hspost}@{}}%
\column{8E}{@{}l@{}}%
\column{10}{@{}>{\hspre}c<{\hspost}@{}}%
\column{10E}{@{}l@{}}%
\column{11}{@{}>{\hspre}l<{\hspost}@{}}%
\column{13}{@{}>{\hspre}l<{\hspost}@{}}%
\column{24}{@{}>{\hspre}l<{\hspost}@{}}%
\column{E}{@{}>{\hspre}l<{\hspost}@{}}%
\>[5]{}\Varid{q}{}\<[8]%
\>[8]{}\mathbin{:}{}\<[8E]%
\>[11]{}\Conid{T}\mathbin{→}\Conid{Q}{}\<[E]%
\\
\>[5]{}\Varid{q}\;\Varid{t}{}\<[10]%
\>[10]{}\mathrel{=}{}\<[10E]%
\>[13]{}\Varid{w}\;\Varid{t}{}\<[24]%
\>[24]{}\mbox{\onelinecomment  or, equivalently, \ensuremath{\Varid{q}\mathrel{=}\Varid{w}}}{}\<[E]%
\\[\blanklineskip]%
\>[5]{}\dot{q}\mathbin{:}\Conid{T}\mathbin{→}\Conid{V}{}\<[E]%
\\
\>[5]{}\dot{q}\;\Varid{t}\mathrel{=}\Varid{dw}\mathbin{/}\Varid{dt}{}\<[24]%
\>[24]{}\mbox{\onelinecomment  or, equivalently, \ensuremath{\dot{q}\mathrel{=}\Conid{D}\;\Varid{w}}}{}\<[E]%
\ColumnHook
\end{hscode}\resethooks
  We combine these in the ``combinator'' \ensuremath{\Varid{expand}}, given by
  \begin{hscode}\SaveRestoreHook
\column{B}{@{}>{\hspre}l<{\hspost}@{}}%
\column{5}{@{}>{\hspre}l<{\hspost}@{}}%
\column{17}{@{}>{\hspre}c<{\hspost}@{}}%
\column{17E}{@{}l@{}}%
\column{20}{@{}>{\hspre}l<{\hspost}@{}}%
\column{E}{@{}>{\hspre}l<{\hspost}@{}}%
\>[5]{}\Varid{expand}\mathbin{:}(\Conid{T}\mathbin{→}\Conid{Q})\mathbin{→}(\Conid{T}\mathbin{→}(\Conid{T},\Conid{Q},\Conid{V})){}\<[E]%
\\
\>[5]{}\Varid{expand}\;\Varid{w}\;\Varid{t}{}\<[17]%
\>[17]{}\mathrel{=}{}\<[17E]%
\>[20]{}(\Varid{t},\Varid{w}\;\Varid{t},\Conid{D}\;\Varid{w}\;\Varid{t}){}\<[E]%
\ColumnHook
\end{hscode}\resethooks

\item With \ensuremath{\Varid{expand}} in our toolbox we can fix the typing problem in
  item \ref{item:L:contra} above.
  The Lagrangian is a ``function of the system state (time,
  coordinates, and velocities)'' and the ``expanded path'' (\ensuremath{\Varid{expand}\;\Varid{w}}) computes the state from just the time.
  By composing them we get a function
  \begin{hscode}\SaveRestoreHook
\column{B}{@{}>{\hspre}l<{\hspost}@{}}%
\column{5}{@{}>{\hspre}l<{\hspost}@{}}%
\column{21}{@{}>{\hspre}c<{\hspost}@{}}%
\column{21E}{@{}l@{}}%
\column{24}{@{}>{\hspre}l<{\hspost}@{}}%
\column{E}{@{}>{\hspre}l<{\hspost}@{}}%
\>[5]{}\Conid{L}\mathbin{\circ}(\Varid{expand}\;\Varid{w}){}\<[21]%
\>[21]{}\mathbin{:}{}\<[21E]%
\>[24]{}\Conid{T}\to \Conid{ℝ}{}\<[E]%
\ColumnHook
\end{hscode}\resethooks
  which describes how the Lagrangian would vary over time if the
  system would evolve according to the path \ensuremath{\Varid{w}}.

  This particular composition is not used in the equation, but we do
  have
  \begin{hscode}\SaveRestoreHook
\column{B}{@{}>{\hspre}l<{\hspost}@{}}%
\column{5}{@{}>{\hspre}l<{\hspost}@{}}%
\column{32}{@{}>{\hspre}c<{\hspost}@{}}%
\column{32E}{@{}l@{}}%
\column{35}{@{}>{\hspre}l<{\hspost}@{}}%
\column{E}{@{}>{\hspre}l<{\hspost}@{}}%
\>[5]{}(\mathbin{∂}\Conid{L}\mathbin{/}\mathbin{∂}\dot{q})\mathbin{\circ}(\Varid{expand}\;\Varid{w}){}\<[32]%
\>[32]{}\mathbin{:}{}\<[32E]%
\>[35]{}\Conid{T}\to \Conid{ℝ}{}\<[E]%
\ColumnHook
\end{hscode}\resethooks
  which is used inside \ensuremath{\Varid{d}\mathbin{/}\Varid{dt}}.

\item We now move to using \ensuremath{\Conid{D}} for \ensuremath{\Varid{d}\mathbin{/}\Varid{dt}}, \ensuremath{\Conid{D₂}} for \ensuremath{\mathbin{∂}\mathbin{/}\mathbin{∂}\Varid{q}}, and
  \ensuremath{\Conid{D₃}} for \ensuremath{\mathbin{∂}\mathbin{/}\mathbin{∂}\dot{q}}.
  In combination with \ensuremath{\Varid{expand}\;\Varid{w}} we find these type correct
  combinations for the two terms in the equation:
  \begin{hscode}\SaveRestoreHook
\column{B}{@{}>{\hspre}l<{\hspost}@{}}%
\column{5}{@{}>{\hspre}l<{\hspost}@{}}%
\column{8}{@{}>{\hspre}l<{\hspost}@{}}%
\column{16}{@{}>{\hspre}c<{\hspost}@{}}%
\column{16E}{@{}l@{}}%
\column{19}{@{}>{\hspre}l<{\hspost}@{}}%
\column{32}{@{}>{\hspre}c<{\hspost}@{}}%
\column{32E}{@{}l@{}}%
\column{35}{@{}>{\hspre}l<{\hspost}@{}}%
\column{E}{@{}>{\hspre}l<{\hspost}@{}}%
\>[5]{}\Conid{D}\;((\Conid{D₃}\;\Conid{L}){}\<[16]%
\>[16]{}\mathbin{∘}{}\<[16E]%
\>[19]{}(\Varid{expand}\;\Varid{w})){}\<[32]%
\>[32]{}\mathbin{:}{}\<[32E]%
\>[35]{}\Conid{T}\mathbin{→}\Conid{ℝ}{}\<[E]%
\\
\>[5]{}\hsindent{3}{}\<[8]%
\>[8]{}(\Conid{D₂}\;\Conid{L}){}\<[16]%
\>[16]{}\mathbin{∘}{}\<[16E]%
\>[19]{}(\Varid{expand}\;\Varid{w}){}\<[32]%
\>[32]{}\mathbin{:}{}\<[32E]%
\>[35]{}\Conid{T}\mathbin{→}\Conid{ℝ}{}\<[E]%
\ColumnHook
\end{hscode}\resethooks

  The equation becomes
  \begin{hscode}\SaveRestoreHook
\column{B}{@{}>{\hspre}l<{\hspost}@{}}%
\column{5}{@{}>{\hspre}l<{\hspost}@{}}%
\column{30}{@{}>{\hspre}c<{\hspost}@{}}%
\column{30E}{@{}l@{}}%
\column{33}{@{}>{\hspre}l<{\hspost}@{}}%
\column{54}{@{}>{\hspre}c<{\hspost}@{}}%
\column{54E}{@{}l@{}}%
\column{57}{@{}>{\hspre}l<{\hspost}@{}}%
\column{E}{@{}>{\hspre}l<{\hspost}@{}}%
\>[5]{}\Conid{D}\;((\Conid{D₃}\;\Conid{L})\mathbin{∘}(\Varid{expand}\;\Varid{w})){}\<[30]%
\>[30]{}\mathbin{-}{}\<[30E]%
\>[33]{}(\Conid{D₂}\;\Conid{L})\mathbin{∘}(\Varid{expand}\;\Varid{w}){}\<[54]%
\>[54]{}\mathrel{=}{}\<[54E]%
\>[57]{}\Varid{const}\;\mathrm{0}{}\<[E]%
\ColumnHook
\end{hscode}\resethooks
  or, after simplification:
  \begin{hscode}\SaveRestoreHook
\column{B}{@{}>{\hspre}l<{\hspost}@{}}%
\column{5}{@{}>{\hspre}l<{\hspost}@{}}%
\column{26}{@{}>{\hspre}c<{\hspost}@{}}%
\column{26E}{@{}l@{}}%
\column{29}{@{}>{\hspre}l<{\hspost}@{}}%
\column{E}{@{}>{\hspre}l<{\hspost}@{}}%
\>[5]{}\Conid{D}\;(\Conid{D₃}\;\Conid{L}\mathbin{∘}\Varid{expand}\;\Varid{w}){}\<[26]%
\>[26]{}\mathrel{=}{}\<[26E]%
\>[29]{}\Conid{D₂}\;\Conid{L}\mathbin{∘}\Varid{expand}\;\Varid{w}{}\<[E]%
\ColumnHook
\end{hscode}\resethooks
  where both sides are functions of type \ensuremath{\Conid{T}\mathbin{→}\Conid{ℝ}}.

\item ``A path is allowed if and only if it satisfies the Lagrange
  equations'' means that this equation is a predicate on paths (for a
  particular \ensuremath{\Conid{L}}):
  \begin{hscode}\SaveRestoreHook
\column{B}{@{}>{\hspre}l<{\hspost}@{}}%
\column{5}{@{}>{\hspre}l<{\hspost}@{}}%
\column{23}{@{}>{\hspre}l<{\hspost}@{}}%
\column{E}{@{}>{\hspre}l<{\hspost}@{}}%
\>[5]{}\Conid{Lagrange}\;(\Conid{L},\Varid{w})\mathrel{=}{}\<[23]%
\>[23]{}\qquad\Conid{D}\;(\Conid{D₃}\;\Conid{L}\mathbin{∘}\Varid{expand}\;\Varid{w})\doubleequals\Conid{D₂}\;\Conid{L}\mathbin{∘}\Varid{expand}\;\Varid{w}{}\<[E]%
\ColumnHook
\end{hscode}\resethooks
  where we use \ensuremath{(\doubleequals)} to avoid confusion with the equality sign (\ensuremath{\mathrel{=}})
  used for the definition of the predicate.
\end{enumerate}

So, we have figured out what the equation ``means'', in terms of
operators we recognise.
If we zoom out slightly we see that the quoted text means something
like:
If we can describe the mechanical system in terms of ``a Lagrangian''
(\ensuremath{\Conid{L}\mathbin{:}\Conid{S}\to \Conid{ℝ}}), then we can use the equation to check if a particular
candidate path \ensuremath{\Varid{w}\mathbin{:}\Conid{T}\mathbin{→}\Conid{ℝ}} qualifies as a ``motion of the system'' or
not.
The unknown of the equation is the path \ensuremath{\Varid{w}}, and as the equation
involves partial derivatives it is an example of a partial
differential equation (a PDE).

It is instructive to compare our treatment to Sussman and Wisdom's
own answer to their question.
Although their approach is very similar in spirit (``we use computer
programming in a functional style to encourage clear thinking'',
\cite[page \emph{xv}]{sussman2013functional}), we and our students
have found their explanations hard to follow.
Their main vehicle for expressing computations, the programming
language Scheme, is dynamically typed, and encourages a style in which
the typing information is left implicit.
As this example illustrates,
we have found that having explicit typing is the better choice.

\section{Evaluation and results}

Now we turn to the empirical evaluation of the course in terms of the
students' pass rates and grades in related courses.
We considered student results for students from the CSE programme at Chalmers%
\footnote{Computer Science and Engineering (CSE) is a five-year BSc+MSc
  programme at Chalmers. It is called ``Datateknik'' (abbreviated ``D'') in Swedish.}
who started their studies in 2014 and 2015. In the spring of their second year at
Chalmers (2016 and 2017), these students had the option of either taking the
DSLsofMath course or a course on Concurrent Programming (see Table~\ref{tab:CSE-prog}).

\begin{table}[htb]
  \centering
  \begin{tabular}{lll}
                       & Fall  & Spring\\
    \hline
    Year 1  & \textcolor{gray}{Compulsory courses}  & \textcolor{gray}{Compulsory courses} \\
    Year 2  & \textcolor{gray}{Compulsory courses}  & \textcolor{red}{DSLsofMath} OR ConcProg \\
    Year 3  & \textcolor{blue}{TSS + Control}  & ... \\

  \end{tabular}
\caption{CSE programme structure (simplified)}
\label{tab:CSE-prog}
\end{table}

We considered only ``active'' students, that is, students who had signed up for
at least half of the compulsory courses in the CSE programme during the semesters
being considered (Fall 2014 - Fall 2017). This amounted to 145 students, where
53 signed up for the DSLsofMath course (whereas 92 did not) and 34 of those 53
passed the course.

We had access to data on these students' results in all compulsory courses in
the programme as well as in the more common elective courses (21 courses in all).
At Chalmers students pass a course with a grade of 3, 4, or 5, with 5
being the highest grade, or fail the course with no specified grade.
Students usually have two yearly opportunities to retake exams from
courses they took in past semesters, to attempt to obtain a passing
score or to improve their grades.

The enrolment and results of the DSLsofMath course itself was as follows:
\begin{itemize}
\item 2016: 28 students, pass rate: 68\%
\item 2017: 43 students, pass rate: 58\%
\item 2018: 39 students, pass rate: 89\%
\end{itemize}
Note that this also counts students from other programmes (mainly SE
and Math) while the rest only deals with the CSE students.

\subsection{Student results before and after taking DSLsofMath}

Our hope was that taking our course would help prepare the students
for the math-intensive compulsory courses in the subsequent year with which
many students struggle;
Transforms, signals and systems (TSS%
\footnote{In Swedish:
  \href{https://student.portal.chalmers.se/en/chalmersstudies/courseinformation/Pages/SearchCourse.aspx?course_id=28314&parsergrp=3}{SSY080
    Transformer, signaler och system}})
and Control theory (Control%
\footnote{In Swedish:
  \href{https://student.portal.chalmers.se/en/chalmersstudies/courseinformation/Pages/SearchCourse.aspx?course_id=24149&parsergrp=3}{ERE103
    Reglerteknik}})

In Table~\ref{tab:passratesafter} we see the pass rate and the mean
grade (of those who passed) for the above mentioned courses, where
PASS represents the students who took the DSLsofMath course and passed
it, IN is the group of students who took DSLsofMath (whether or not
they passed), and OUT is the students who did not sign up for
DSLsofMath.
\begin{table}[htb]
  \centering
  \begin{tabular}{l*{3}{c}}
                       & PASS  & IN   & OUT  \\
    \hline
    TSS pass rate   & 77\%  & 57\% & 36\% \\
    TSS mean grade  & 4.23  & 4.10 & 3.58 \\
    Control pass rate   & 68\%  & 45\% & 40\% \\
    Control mean grade  & 3.91  & 3.88 & 3.35 \\

  \end{tabular}
  \caption{Pass rate and mean grade in third year courses for students who took and
  passed DSLsofMath and those who did not.}
\label{tab:passratesafter}
\end{table}

As we can see, the students who took DSLsofMath had higher mean grades in the
third-year courses and were more likely to pass, in particular those who managed
to pass the DSLsofMath course.
The correlation between taking, and especially passing, DSLsofMath and
success in the third year courses is clear.
But perhaps the students who chose to take our course did so because
they enjoyed mathematics, and were already more likely to succeed in
the subsequent math-heavy courses regardless of whether they took our
course.
%\subsubsection{Characterising students who chose DSLsofMath}
%We were curious to see how we could characterize the students who sign up for
%our course, for instance whether they were likely to be students who had done
%well in mathematics in the past who were interested in studying more maths,
%or perhaps students who had struggled with mathematics in the past and were
%therefore looking for more support and new methods for studying maths.
To explore this, we looked at the students' results from their first
three semesters at Chalmers, prior to having the option of taking
DSLsofMath.
We were particularly interested in students' performance in the
compulsory mathematics and physics courses (see
Table~\ref{tab:passratesbefore}, and appendix \ref{app:coursecodes}).
\begin{table}[htb]
  \centering
  \begin{tabular}{l*{3}{c}}
                                     & PASS  & IN   & OUT  \\
    \hline
    Pass rate for first 3 semesters  & 97\%  & 92\% & 86\% \\
    Mean grade for first 3 semesters & 3.95  & 3.81 & 3.50 \\
    Math/physics pass rate           & 96\%  & 91\% & 83\% \\
    Math/physics mean grade          & 4.01  & 3.84 & 3.55 \\

  \end{tabular}
  \caption{Pass rate and mean grade for courses taken prior to taking (or not
    taking) DSLsofMath.}
\label{tab:passratesbefore}
\end{table}

Here we can see that there seems to be a small positive bias: those
students who choose the DSLsofMath course (the IN group) were a bit
more successful in the first three semesters than those in the OUT
group.
And (not surprisingly) those who pass the course fare even better, on
average.
Given that many other factors also vary it is not easy to prove that
taking DSLsofMath is a significant factor in improving future results,
but it does seem likely.

% TODO Perhaps include the summary table:
% \begin{table}[h]
%   \centering
%   \begin{tabular}{l*{3}{c}}
%                                      & PASS  & IN   & OUT  \\
%     \hline
%     Pass rate for first 3 semesters  & 97\%  & 92\% & 86\% \\
%     Math/physics pass rate           & 96\%  & 91\% & 83\% \\
%     TSS pass rate                    & 77\%  & 57\% & 36\% \\
%     Control pass rate                & 68\%  & 45\% & 40\% \\
%
%   \end{tabular}
%   %\caption{Pass rate and mean grade for courses taken prior to taking (or not
%   % taking) DSLsofMath.}
% \end{table}
%   Group sizes: PASS 34, IN 53, OUT 92 (145 in all)
%

\subsection{Students' course assessment and resulting changes}

Each course instance has been evaluated with a standard questionnaire sent to
all participants as part of the university-wide course evaluation process.
The evaluation of the first instance, with Cezar Ionescu as lecturer, identified
a need to restructure the initial four-lecture sequence, to re-order a few
lectures and to replace the two guest lectures by Linear Algebra.

In preparation for the second instance, with Patrik Jansson as lecturer, the
initial lecture sequence was changed to include more Haskell introduction and
less formal logic, and two new lectures on Linear Algebra were developed.
In the evaluation of the second (2017) instance, the students requested more
lecture notes and more weekly exercises to make it easier to get started.
The evaluation also indicated that the average student had spent too few hours
on the course, and the exam results suffered (42\% failed, up from 32\% in 2016).

At this point we decided to push for course material development and one of the
student evaluators from 2017 (Daniel Heurlin) was hired part time to help out
with these improvements.
Patrik spent the autumn of 2017 on converting raw text notes and photos of
blackboards to LaTeX-based literate Haskell lecture notes covering the full
course and Daniel developed more exercises to solve.
The primary focus was on complementing the exam questions from earlier years
with easier exercises to start each week with and with more material on
functional programming in Haskell.

The recent evaluation of the 2018 instance was overall positive and the course
saw a strong improvement in the pass rate: only 11\% failed.
The student evaluators suggested to ``increase the pressure'' on solving the
exercises, to make the students better prepared for the hand-in assignments and
the written exam.

\section{Related work}

Others have worked on using functional programming to help teach mathematics, in
particular algebra, to younger (primary and secondary school) students.
The Bootstrap project has
successfully developed a functional programming-based curriculum that has
improved students' performance in solving algebra
problems~\cite{Schanzer:2018:ABA:3159450.3159498,Schanzer:2015:TSS:2676723.2677238}.
In \cite{EPTCS270.2}, d'Alves et al.\ describe using functional programming in
Elm to introduce algebraic thinking to students.

In \cite{DBLP:journals/corr/Walck16} and \cite{DBLP:journals/corr/Walck14},
Walck describes using Haskell programming to deepen university students'
understanding of physics, and in \cite{EPTCS106.3} Ragde describes using
functional programming to introduce university students to more precise
mathematical notation.
We are not aware of previous literature on the use of functional programming to
present mathematical analysis as we have done.

\section{Conclusions and future work}

During the last four years we have developed course material and worked with
150+ computer science students to improve their mathematical education through
the course ``Domain Specific Languages of Mathematics''.
We have shown how mathematical concepts like limits, derivatives and Lagrangian
equations can be explored and explained using typed functional programming.
% TODO perhaps include for the full paper Sometimes new insights arise: Stream calculus, for example.
%
(Much more about that can be read in the lecture notes
\cite{JanssonIonescuDSLsofMathCourse}.)
We have investigated the group of students who picked DSLsofMath as an elective
and we have measured positive results on later courses with mathematical content.

There are several avenues for future work: upstream and downstream curriculum
changes, better tool support, and empirical evaluation.
\begin{itemize}
\item Upstream, we would really like to work with the teachers of the
  mathematics courses in the first year to see if some of the ideas from
  DSLsofMath could be included already at that stage.
  Ideally, in the long term, the DSLsofMath course material should be
  ``absorbed'' by these earlier courses.
\item Downstream, it would be interesting to see how the new course could affect
  the way the ``Transforms, signals and systems'' and ``Control theory''
  courses are taught.
  It seems that we could also affect the Physics course -- see the BSc project
  ``Learn You a Physics'' summary in appendix \ref{app:LearnYouAPhysics}.
\item When it comes to tool support it would be interesting to see how proof systems
  like Liquid Haskell, Agda, etc.\ could help the students learn.
  We have been cautious so far, taking it one step at a time, to avoid stressing
  the student with yet another language / system / tool to learn.
\item Finally, we are well aware that our evaluation of the effect of the course
  on the students' learning is lacking the rigour of a proper empirical study.
  It would be interesting to work with experts on teaching and learning in
  higher education on such a study.

\end{itemize}

\subsection*{Acknowledgements}

The support from Chalmers Quality Funding 2015 (Dnr C 2014-1712, based
on Swedish Higher Education Authority evaluation results) is
gratefully acknowledged.
Thanks also to R. Johansson (as Head of Programme in CSE) and P.
Ljunglöf (as Vice Head of the CSE Department for BSc and MSc
education) who provided continued financial support when the national
political winds changed.
Thanks to D. Heurlin who provided many helpful comments during his
work as a student research assistant in 2017.

This work was partially supported by the projects GRACeFUL (grant
agreement No 640954) and CoeGSS (grant agreement No 676547), which
have received funding from the European Union’s Horizon 2020 research
and innovation programme.

\appendix

\section{DSLsofMath learning outcomes}
\label{app:learningoutcomes}

\begin{itemize}
\item Knowledge and understanding
  \begin{itemize}
  \item design and implement a DSL for a new domain
  \item organize areas of mathematics in DSL terms
  \item explain main concepts of analysis, algebra, and lin.\ alg.
  \end{itemize}
\item Skills and abilities
  \begin{itemize}
  \item develop adequate notation for mathematical concepts
  \item perform calculational proofs
  \item use power series for solving differential equations
  \item use Laplace transforms for solving differential equations
  \end{itemize}
\item Judgement and approach
  \begin{itemize}
  \item discuss and compare different software implementations of
        mathematical concepts
  \end{itemize}
\end{itemize}

\url{https://github.com/DSLsofMath/DSLsofMath/blob/master/Course2018.md}

\section{BSc project ``Learn You a Physics''}
\label{app:LearnYouAPhysics}

The online learning material
\href{https://dslsofmath.github.io/BScProj2018/}{``Learn You a Physics''} (by E.
Sjöström, O. Lundström, J. Johansson, B. Werner) is the result of a BSc project
at Chalmers (supervised by P. Jansson) where the goal is to create an
introductory learning material for physics aimed at programmers with a basic
understanding of Haskell.
It does this by identifying key areas in physics with a well defined scope, for
example dimensional analysis or single particle mechanics, and develops a domain
specific language around each of these areas.
The implementation of these DSLs are the meat of the learning material with
accompanying text to explain every step and how it relates to the physics of
that specific area.
The text is written in such a way as to be as non-frightening as possible, and
to only require a beginner knowledge in Haskell.
Inspiration is taken from \href{http://learnyouahaskell.com/}{Learn You a
  Haskell for Great Good} and the project
\href{https://github.com/DSLsofMath/DSLsofMath}{DSLsofMath} at Chalmers and
University of Gothenburg.
The \href{https://github.com/DSLsofMath/BScProj2018/tree/master/Physics}{source
  code} and \href{https://dslsofmath.github.io/BScProj2018/}{learning material}
is freely available online.

\section{Course codes}
\label{app:coursecodes}
The courses used to calculate the first 3 semesters pass rate and mean
grade are the ones listed above DSLsofMath (DAT326) in Table~\ref{tab:coursecodes}.
Of these, the ones used to calculate the Math/physics pass rate and
mean grade were TMV210, TMV216, TMV170, MVE055 and TIF085.
\begin{table}[htb]
  \centering
\begin{tabular}{lll}
    Course code & & Name
\\\hline
         TDA555 &   & Introduction to functional programming            % Introduktion till funktionell programmering
\\       TMV210 & M & Introduction to discrete mathematics		% Inledande diskret matematik
\\       EDA452 &   & Introduction to computer engineering		% Grundläggande datorteknik
\\       TMV216 & M & Linear algebra					% Linjär algebra
\\       DAT043 &   & Object oriented programming			% Objektorienterad programmering
\\       TMV170 & M & Calculus						% Matematisk analys
\\       EDA343 &   & Computer communication				% Datakommunikation
\\       EDA481 &   & Machine oriented programming			% Maskinorienterad programmering
\\       DAT290 &   & Computer science and engineering project		% Datatekniskt projekt
\\       MVE055 & M & Mathematical statistics and discrete mathematics	% Matematisk statistik och diskret matematik
\\       DAT037 &   & Data structures					% Datastrukturer
\\       TIF085 & F & Physics for engineers				% Fysik för ingenjörer
\\\hline DAT326 & * & Domain Specific Languages of Mathematics		% Matematikens domänspecifika språk
\\\hline SSY080 &   & Transforms, signals and systems			% Transformer, signaler och system
\\       ERE103 &   & Control theory					% Reglerteknik
\end{tabular}
\caption{Course codes}
\label{tab:coursecodes}
\end{table}

\bibliographystyle{eptcs}
\bibliography{dslm}
\end{document}